\documentstyle{article}
\begin{document}
\author{Eugene I. Shtyrkov}

\title{The Evolved-Vacuum Model of  Redshifts}
\frenchspacing

\maketitle
\begin{abstract}

A new interpretation of cosmological redshifts is proposed to construct the
evolved-vacuum model of this phenomenon.The physical vacuum was considered
 to be a real matter with time-dependent  permittivity and permeability.Time
variation of these parameters ($\frac{\Delta
\varepsilon}{\varepsilon}=\frac{\Delta \mu}{\mu}=9\cdot 10^{-11}$ per
year)  was
shown on the base of Maxwell's equations and
Hubble's law to obey the exponential behavior causing step-by-step decrease
of light velocity with a rate of about 2.7 m/s over 100 years.Cosmological
 aspects are discussed to explain some features of reality of the Universe
evolution.
\end{abstract}

\section{Introduction}

It was experimentally established early in this century that the fainter
 distant galaxies and quasars are, the larger their shift of spectral lines
toward the red region is. Assuming a plausible faintness-distance function,
Hubble~\cite{hu} discovered that there is a linear dependence between a
redshift  and distance  $L$
\begin{equation}
Z(L)=\frac{\lambda_{shift}-\lambda_{o}}{\lambda_{o}}=\left( \frac{H_{o}}{c_{o}} \right) L
\end{equation}
In this relation, $Z(L)$  is the relative spectral shift,$\lambda_{o}$   is
the wavelength of a spectral line from a source in the laboratory which is at
 rest relative to the Earth, $\lambda_{shift}$  is the wavelength of the same
line emitted by a quasar and  measured by a terrestrial observer in the
laboratory, $H_{o}$  is  the Hubble constant and  $c_{o}$  is the free space
velocity of light. Until now, such spectral
observations yielding information about extremely distant objects were being the
only experimental  clues to support an understanding of the Universe.
This is because only such gigantic distances  (of about $10^{22}$ km)  and
time intervals  ($10^{9}$ years)  suffice to reveal the small changes that occur
while light is traveling through space.  Therefore, a  {\em correct}  interpretation of such
spectral observations to ascertain the real origin of redshifts and bring the Universe
to light is especially important.

At present, there are several alternative models to explain the redshift phenomenon.
Usually, this one is interpreted as a Doppler effect which finally implies recession
of galaxies and expansion of the Universe (the "Big Bang" model).There are
some doubts, however, about this interpretation .Really, there is no physical
explanation for singularities due to infinite density of matter at the point of
creation, and some recession velocities appear to be extremely large.
For example, for  quasar Q01442+101 (at $Z=3,3$) the recession velocity derived
from this model is
about of  $0.9c_{o}$ . Moreover, some super-clusters of galaxies seem to be older than
the age of the Universe derived from this model. It was experimentally
established recently that very extended objects (the huge sheets of galaxies
stretching more than 200 million l.y. (light-years) across , 700 million l.y. long
and  20 million l.y. thick ) are like  the super-clusters of  galaxies mapped by
Tully earlier \cite{tul} .Still larger objects ({\em "Cosmic Ladder"}) stretching across a distance
of  about seven billion light-years have been discovered \cite{ler}. Because the maximal
velocities of any objects experimentally observed in astronomy to be not more
than  500 km/s  it takes about 150 billion years to form this structure --  more
than seven times  the number of years since the Big Bang to form the Universe \cite{ler}.

Besides the Big-Bang there are also alternative approaches  based on  ideas of
time evolution of matter/light parameters, either the fundamental physical
constants (Plank's constant, charge and mass of elementary particles, the
electromagnetic parameter $c_{o}$ , and so on) or the electromagnetic
characteristics of the physical vacuum.\\
The variation of fundamental physical constants as a possible origin of the
redshifts has been discussed widely since Dirac \cite{di} put forward the idea of
"Great Numbers". Recently, however, some of the atomic constants  was really
shown to be constant  in time.For instance,Potekhin and Varshalovich \cite{PV} have studied the fine
splitting of the doublet absorption lines in quasar's spectra.  They analyzed
1414 doublets (CIV, NV, CVI, MgII, AlIII, and SiIV) in the wide range of redshifts
 $ (0.2 < Z< 3.7)$.  Their statistical analysis reveals no statistically significant
 time variation of  the fine structure constant  $\alpha=\frac{e^{2}}{hc_{o}}$    on
 time scale of about ten billion years.We can conclude from here that, at least,
 such the parameters as charge of electron $e$,
 Plank's constant $h$ and electromagnetic coefficient  $c_{o}$  should be considered
 as constant ones and, hence, can have no influence on redshifts. As for variation
  of the mass of elementary particles, this idea discussed by Arp \cite{ar} in the
  intrinsic-redshift model to prove that redshifts are supposedly related to the age
   of the objects. This idea, however, do not obey the redshift-distance relation (1)
   at constant $H_{o}$ \cite{arp} . Moreover, constancy of  the electron mass in time
    appears to follow from the results of \cite{PV}  as well.Really,because the fine splitting of energy
    levels $\delta E$  depends on
    Ridberg's  constant  $R=\frac{2\pi^{2} m e^{4}}{c_{o} h^{3}}$  as well
    we should conclude that $m$  is also {\em constant on a very large time scale}.

There are also alternative  models of redshifts which obey the redshift-distance
relation and based on an idea of gradual change of light parameters due to
interaction between light and matter while the light is traveling gigantic
distances through space for a very long time. There are two candidate ways
for such interaction to cause redshifts: gradual energy loss by the photon
due to absorption during propagation of light with a constant velocity
(tired-light model,see,for instance, \cite{DeB}) and  propagation of light with
the variable velocity and without  absorption  in free space (variable-light-velocity models).\\
Tired-light mechanism, however, results in obvious contradictions between
quantum and classical description based on the Maxwell's equations.
In fact, assuming that energy of a single photon is gradually decreasing due
to absorption, we may conclude that volume energy density of  N photon
flow with this same frequency and,hence,its intensity are also decreasing ones.
From the electrodynamics point of view, it means that electric field strength
 $E$ should gradually be decreased  while this wave is traveling through space .
Quantum description says about simultaneous decreasing of frequency at changing
of photon energy.However,it is not difficult to be convinced that no such a
combination, i.e. the simultaneous
decaying space-dependent functions $E(x)$  and  $\omega (x)$ , obeys
 this same wave equation with stationary boundary conditions
which is currently used in quantum electronics and physical optics to
adequately describe the propagation of light at {\em constant}
 velocity in any non-conductors,including vacuum (see Appendix) .
The electromagnetic coefficient $c_o$   , which bridges the electric and magnetic
 phenomena, has the dimension of speed, and has been {\em historically}
  identified with a constant free space light velocity. In any media, however, the
  light velocity depends on its  permittivity  $ \varepsilon $    and permeability
 $ \mu $ as well. At present, vacuum has been experimentally established to be not
  a void but it is some material medium with definite but not so far investigated
  features. It was really confirmed by observation of several vacuum effects, for
   instance, zero oscillations and polarization of vacuum, generating the particles in
   vacuum  due to electromagnetic interaction. Therefore, it was reasonable to
   assume that this real matter-physical vacuum  can possess internal  friction due to
   its small but a real viscosity to result in variation of light-matter interaction.
   That is, vacuum can affect on the light wave because of certain resistance.
   Because physical vacuum is a real material with real characteristics the light
 velocity can be non-constant, since it depends both on $ c_o $ and $\varepsilon$,
$\mu$  of the vacuum ,all of which could have been space/time-dependent functions,
  in principle. This may be a reason for the redshifts observed.
   For example, the situation with space variable $\varepsilon(x)$  and
    $\mu(x)$  at  constant  electromagnetic parameter was discussed in \cite{S}
     where a wave equation  with   the term analogous to one for a damped simple oscillator was derived from
 Maxwell's equations. Solution of this equation leads to a gradual  increase
  of a  wavelength ( redshifts at constancy of the  frequency) and variable light
   velocity.
   A drawback of  the last  approach is what one need to consider  the permittivity
and permeability rather as  parameters of interaction but not just the characteristics
 of vacuum as real matter. This inconvenience was overcame in \cite{Sh} where this
  same result was obtained, but at constant permittivity/permeability and the
   space-dependent electromagnetic parameter $c_o$  . However, following the
    publication of the work \cite{PV}  followed by conclusion about constancy of
     $c_o$    the last  model  \cite{Sh}  must be
obviously abandoned .
In the present paper we will consider more promising  variable light velocity
 model of redshifts which let us join electrodynamical approach with the
 cosmological principle and empirical data to be available in order to explain some features of reality.
 \section{Evolved-vacuum model (EVM) of cosmological redshifts}
 This model is based on classical electrodynamics with taking the time-dependent
 permittivity and permeability into account \cite{Sht}.
 Let us make only one assumption that, in compliance with the cosmological
 principle, the variation of the physical vacuum parameters occurs simultaneously
 and identically {\em at any point \/} of the  infinite evolving Universe. Then the
 permittivity and permeability of the physical vacuum at the moment when light
 is leaving a distant galaxy (one point of space) would be different from what
  it would be  when this light is reaching  the Earth  (the other point of the
   Universe) to be a reason for shifts of spectral lines.
Let us consider this point in more detail by writing Maxwell's equations for
the plane-polarized monochromatic wave propagating along the OX-axis and
$\varepsilon(t)$ and $\mu(t)$  as the functions of time

\begin{eqnarray}
  \frac{\partial E}{\partial x} &=& - \frac{1}{c_o} \frac{\partial B}{\partial t}  \\
  \frac{\partial H}{\partial x} &=& - \frac{1}{c_o} \frac{\partial D}{\partial t} \nonumber \\
   B &=& \mu(t) H \nonumber \\
   D &=& \varepsilon (t) E \nonumber
  \end{eqnarray}

Consideration of the light wave as a plane one here is due to  quasars
removed on infinity are practically  point sources of  light with  flat
wave fronts near the Earth.
The simplest way of analyzing the situation under these conditions is to
write down a wave equation for the induction wave $D$   instead of the
electric field strength $E$. One can argue that wave characteristics of
induction have the same phase behavior as for the electric field.
 However, solving the induction wave equation and then making use of the
 material relations in (2) to find the electric field strenth is much simpler.
 The wave equation for induction can be derived from a chain of substitutions \\
 \begin{displaymath}
 \varepsilon \frac{\partial}{\partial x} \left( \frac{\partial E}{\partial x} \right) = -\frac{\varepsilon}{c_o} \frac{\partial^2 B}{\partial x \partial t}= \frac{\varepsilon}{c_{o}^{2}} \frac{\partial}{\partial t} \left( \mu \frac{\partial D}{\partial t} \right)
 \end{displaymath}
 drawn from Eqs.2 by means of taking a partial derivative of the left part of
 the first equation with respect to $x$  and using the second equation in (2).
 This leads to the wave equation for electric induction
 \begin{equation}
 \frac{\partial^2 D}{\partial x^2}  - \frac{\varepsilon \mu}{c_o ^2} \left(
\frac{\partial^2 D}{\partial t^2} + \frac{1}{\mu} \frac{\partial \mu}{\partial t}  \frac{\partial D}{\partial t}\right)=0
 \end{equation}
 with the boundary and initial conditions : \\ at $x=0$  and  $t=t_{s}$  ( $t_s$ is the
 start instant when the light left the remote source)  the electric field
  strength in the wave zone is \\ $E(0,t_s)=E_o \exp \imath \omega_o t_s$ ,\\
where the  amplitude  and frequency are constant, and $$D (0,t_s) = \varepsilon (t_s) E(0,t_s)$$
One can see from (3) that the vacuum propagation velocity of the induction
wave is a time-dependent function
\begin{equation}
c(t)=\frac{c_o}{\sqrt{\varepsilon (t) \mu (t)}}
\end{equation}
and this must be the same as the light velocity.
Let us seek a solution of  Eq.3 as a quasi-periodic function with variable phase
\begin{equation}
 D=a \exp \imath  \phi (x,t)
\end{equation}
The induction is an electric field strength in a void,i.e. without
any matter (including physical vacuum) filling a space.Because there is
no field-matter interaction in the void  the amplitude of iduction in (5) can
be considered as a constant.
Differentiating (5), inserting into (3) and separating real and imaginary
parts, we obtain two equations
\begin{eqnarray}
\frac{\partial \phi}{\partial x}=\pm \frac{1}{c} \frac{\partial \phi}{\partial t}   \\
\frac{\partial^2 \phi}{\partial x^2} -q(t) \frac{\partial^2 \phi}{\partial t^2}
- p(t) \frac{\partial \phi}{\partial t} =0 \nonumber
\end{eqnarray}
where  $q(t)=\frac{1}{c^{2}(t)}$ and $p(t)=\frac{q}{\mu (t)} \frac{d \mu (t)}{dt}$.
In order to admit time dependence for both permeability and permittivity, we
should repeat the same analysis for the magnetic induction wave  equation.
Following the differentiating of the  left side of the second equation (2) with
respect to $x$ and making the necessary substitutions using the first one we
obtain the same  equation as (3) but with
$B$ in place of $D$ and $\varepsilon$    in place of $\mu$   in the bracket in the third term.
Obviously, the solution is formally the same as (6), but with $\varepsilon$
in place of $\mu$   in the  definition of $p(t)$.
Using this conclusion, and the  definition of $c$ in (4) , we obtain
\begin{equation}
\frac{dc(t)/dt}{c(t)}=- \frac{d \mu (t)/dt}{\mu (t)}=- \frac{d \varepsilon (t)/dt}{\varepsilon}=Q
\end{equation}
where  $Q$ is either an as-yet unknown function on time or a constant.
Taking it into account  we can rewrite  the Eqs. (6) as follows
\begin{eqnarray}
\frac{\partial \phi}{\partial x}=\pm \frac{1}{c}\frac{\partial \phi}{\partial t}   \\
\frac{\partial^2 \phi}{\partial x^2} - \frac{1}{c^2} \frac{\partial^2 \phi}{\partial t^2} - \left( \frac{1}{c^3} \frac{dc}{dt} \right) \frac{\partial \phi}{\partial t} =0  \nonumber
\end{eqnarray}

It is seen from (7) that the behavior of the light velocity is the  same as
 permeability and permittivity  time behavior  at any point of space
  (in compliance with the cosmological principle as well).This means that an
  observer at any concrete space point on the light path sees the wave as a
  periodic function whose period depends on the light velocity at this epoch.
    In other words, the light frequency perceived by the observer  depends on
     time alone. Thus, the right part of the first equation (8), which keeps
the frequency $(\frac{\partial \phi}{\partial t})$ and light velocity, depends only on time. Therefore, let us
seek the phase of the light in the form $\phi (x,t)=\varsigma (t)\pm \eta (x)$.
Inserting this form into the first equation (8), we derive
\begin{equation}
\frac{d \eta(x)}{dx}=\pm \frac{1}{c}\frac{d \varsigma (t)}{dt} =k
\end{equation}
Since $\eta (x)$ depends only on space and $\varsigma (t)$  depends only on time,
the parameter $k$ must be constant. It follows from solving the Eq.9 that
$\eta (x)$  is a linear function of $x$ ,that is $\eta (x)=\pm kx +\phi_o$ .
 It is easy to show that this form will obey the second of Eqs. (8) as well.
 Thus the parameter $k$ is a spatial derivative of phase, or a spatial
 frequency,i.e.  a well-known wave number $k=\frac{2 \pi}{\lambda}$.Thus we
 have for phase
\begin{equation}
\phi (x,t)=\varsigma (t) \pm kx
\end{equation}
Thus we come to a very important conclusion: the induction wave, and hence the light one, must travel in vacuum with {\em conservation of wave length}
 even when the parameters are time dependent. This wave length  is determined by
 the initial and boundary conditions (at the point of a quasar location at the
moment  of start of the light wave ,i.e. when light is leaving  the quasar)
\begin{equation}
 \lambda (t_s) = \frac{2 \pi}{k}= \frac{2\pi c(t_s)}{\omega (t_s)}=\lambda_{shift}=const
\end{equation}
where $\omega (t_s)=\omega_o$ -  the frequency of the atomic transition in
question,$t_s$ is the start instant, when light left the source, $c(t_s)$ -
the velocity of light at the concrete {\em epoch} of the Universe evolution.
The time dependent frequency $\omega(t)=\partial\phi/\partial t$,then, can be
inferred from (8), given $c(t)$  , which can in turn be inferred from (7), given $Q(t)$.
To determine this, let us refer to the redshift-distance relation (1).
The distance covered by light depends on $t_s$ and $t_o$-the observation time
 (our epoch at the Earth)  and can be written as
\begin{equation}
L(t_{o},t_s)=\int_{t_s}^{t_o} c(t)dt
\end{equation}
Taking  this and (11) into account, we can rewrite the relation (1) in the form
\begin{equation}
Z(t_s ,t_o)=\frac{\lambda (t_s)}{\lambda_{o}} -1=\frac{H_o}{c_o} \int_{t_s}^{t_o} c(t)dt
\end{equation}
where initial conditions for the wave lengths of light are  (11) for a remote
 source (at $t_s$) and  (14)  for a terrestrial source (at our epoch )
\begin{equation}
 \lambda_{o}=\frac{2\pi c_o}{\omega_o}
\end{equation}
 In the relation (13) the light velocity $c_o=c(t_o)$ and $H_o$ ,  measured at
 our epoch $t_o$, should be taken the same for different remote objects
 observed. Hence, the start moment should be an integration variable.
 Inserting (11) and (14) into Eq.13, differentiating it with respect to $t_s$ ,
  using (12) to infer that $ dL(t_o,t_s)/dt_s=-c(t_s)$    and replacing $t_s \rightarrow t$~,
  we obtain a simple differential equation for the light velocity
\begin{equation}
  \frac{dc(t)}{dt}=-H_o c(t)
\end{equation}
This result is in accord with the Eqs.7 derived from the wave equation with
parameter $Q$  set to minus the Hubble's constant $H_o$.Solving these
equations in the range $t_s <t<t_o$ with our initial conditions, we obtain
an exponential law of time variation of the light velocity,  permittivity  and  permeability:
\begin{eqnarray}
c(t)=c(t_s) e^{-H_o(t-t_s)} \\
\varepsilon (t)=\varepsilon (t_s) e^{H_o(t-t_s)} \nonumber \\
\mu (t)=\mu (t_s) e^{H_o(t-t_s)}  \nonumber
\end{eqnarray}
Using (16) in (9) with the initial condition (11) and $k=2\pi/\lambda (t_s)$ ,
we obtain the time-dependent part of the phase for  the time range  of
$t_s<t<t_o$  as follows
\begin{equation}
\varsigma (t)=\frac{\omega_o}{H_o} \left[ 1-e^{-H_o(t-t_s)} \right]
\end{equation}
  Because the time derivative of this phase is the frequency of the light wave
  we obtain the same behavior for frequency as for light velocity at the same time range
  \begin{equation}
  \omega(t)=\omega_o e^{-H_o(t-t_s)}
  \end{equation}
  Thus the induction wave which obeys the wave equation (3) has a constant
  amplitude,the wavelength shifted initially and the variable frequency due to
  gradual time variation of the vacuum parameters equaled throughout the Universe.
  The behavior of the electric field strength can be derived from the last
  material  relation  in (2) with  taking into account (5),(16) and the initial
  conditions in (3) as follows
\begin{equation}
  E(x,t)=E_o e^{-H_o(t-t_s)}  \exp \imath \phi(x,t)
\end{equation}
where the amplitude of the electric field is seen to decline with time,and
the phase is
\begin{equation}
 \phi (x,t)=\frac{\omega_o}{H_o}\left[ 1-e^{-H_o(t-t_s)} \right] - kx
\end{equation}
We may use  expressions (18),(19),(20) for any source  in dependence on situation.
For instance, in order to compare the parameters of  light arrived on the Earth
 with ones measured for terrestrial source at the same instant  $t_o$  (our epoch)
we should put $t=t_o$ in  using  of  (16)  and  (18) .
As a result we have the following.
\begin{description}
\item [remote source] The wave length for the light arrived  from a galaxy
($g-label$) is $\lambda_g(t_o)=\frac{2\pi c(t_o)}{\omega_g(t_o)}$\\
where the light velocity at our epoch is $c(t_o)=c_o$ (from Ex.16 at $t=t_o$ )
and frequency this light perceived by an observer  is
$\omega_g(t_o)=\omega_o\exp \left[ H_o(t_o-t_s) \right]$  (from   Ex.18   at
$t=t_o$ )
\item[terrestrial source] For light from the terrestrial source ($t-label$)
at this same time is $\lambda_t(t_o)=\frac{2\pi c(t_o)}{\omega_t(t_o)}$ where  frequency of
this source is $\omega_t(t_o)=\omega_o$  (from Ex.18 at $t=t_o =t_s$ ,because
there is no time interval between emitting  and observing  the terrestrial
source wave)
\item[comparison] Using  this,  we obtain the relation
$\frac{\lambda_g(t_o)}{\lambda_t(t_o)}=\frac{\omega_o}{\omega_g(t_o)}=e^{H_o \tau_o}$
where $\tau_o=t_o-t_s$.
 Thus, in compliance with experiment, there is the redshift
$$ \lambda_{shift}=\lambda_g(t_o)=\lambda_{o} e^{H_o \tau_o}$$
\end{description}
Although reproducing the conclusions of the tired-light model, namely, about
simultaneous decreasing  the electric field strength and frequency, this model
 has a different physical interpretation. Instead of energy loss due to
absorption at {\em  constant} light velocity,this mechanism  is based on
gradual change of the vacuum parameters that results in declining of the
electric field strength. The electromagnetic wave is gradually slowing down,
with conservation of the initially shifted wavelength $\lambda_{shift}$. The
 frequency perceived by observers at any point on the light path depends on
the light velocity at the observation time.
\section{Cosmological aspects}
The cosmological principle implies that the Eqs.16 derived for the interval
$\tau_o=t_o - t_s$    can be extrapolated from present observation time $t_o$
to any future or past one.If we take our epoch  as zero point on the time scale
the light velocity in the Exp.16  can be rewritten as follows
\begin{equation}
c(t)=c_o e^{-H_o (t-t_o)}
\end{equation}
where $t < t_o$ serves to define history of the Universe before our epoch.
For $t >t_o$  we have future of the Universe.The exponential dependence implies
no particular points or singularities on the time axis.That is, all of the
variations of the Universe parameters have neither beginning nor end but occur
always and everywhere, identically.  Such variation is very small (for instance,
as follows from  Ex.21  for light velocity at $c_o=3 \cdot 10^8 m/s$  it   is  about
of  2.7 m/s for the interval of 100 years).But it is quite measurable with
contemporary techniques. Recently, Montgomery and Dolphin \cite{MD} performed a
statistical analysis of extensive experimental data to argue that light velocity
is variable in time. This analysis shows the measured value of light velocity
to have decreased slightly over the past 250 years. Such behavior of the light
velocity can permit a steady-state cosmology with the boundless Universe that
has always existed, and is homogeneous on the very large scale.
Making use of (12)  with (16), we can find the distance $r$  covered by light
for any moment of time   after the start time   when the light left the quasar
\begin{equation}
 r (t)=\frac{c(t_s)}{H_o}\left[ 1-e^{-H_o(t-t_s)} \right]
\end{equation}
Unlike the constant light-velocity model, this model says that the distance
approaches a certain limit in a certain interval of time $\tau_h=t_h-t_s$ .
At, the $\tau_h ~ \cong (5 \div 6)/H_o$   distance  reaches the limit
$$L_h\cong \frac{c(t_s)}{H_o}$$
This limit distance is due to total declining the electric field strength (19)
and can be interpreted as a spatial cosmological horizon for light.  If we take
this horizon into account the photometric Olbers' paradox \cite{Ol} has a natural explanation.
 Indeed, the light from a galaxy cluster cannot possibly reach the Earth if
the Earth  is situated beyond the light horizon that is available for this
cluster.In other words, a terrestrial observer can see only some remote clusters
the horizons of which are  in excess of the look-back time $\tau_o = t_o-t_s$ ,
 i.e.  for  $\tau_o < \tau_h$  . It is interesting that the earlier light
has been emitted, the larger its horizon is, because of larger light velocity
at this moment of start. Hence the light horizon for quasars associated with
the younger Universe is larger than that for more recent ones. Using  (16)
and (22) at $t=t_o$ and (1)  where $L=L(\tau_o)=r(t_o)$ we obtain the relative
shift
\begin{equation}
 Z(\tau_o)=\frac{H_o}{c_o} L(\tau
_o)=e^{H_o \tau_o} -1
\end{equation}
It follows from evaluation in (23) that the maximal $Z$, being for the horizon
($\tau_h ~ \cong (5 \div 6)/H_o$) is in the range of about 150-500. However,
no empirical $Z$ measured up to now exceed 5. If we take a real declining of
light intensity into account this can be explained in the following way.
In reality, we have a spherical wave front from the point source (remote galaxy)
intensity  of which  is as the inverse square of the distance. In fact, the
relation $$\nabla^2 V=\frac{1}{r} \frac{\partial^2 (rV)}{\partial r^2}$$   is
valid for any function V(r) where the radius of a spherical wave is
$r=\sqrt{ x^2 +y^2+z^2 }$   (see, for instance, \cite{BW}). Therefore, we may place
the product $(rD)$ instead of $D$ into the Eq.3 without changing of it
(at the same directions of the light beam $r$ and $x$).
Thus, we obtain  the induction  in a wave zone ($r >> \lambda/2 \pi$ )
as a spherical wave and,hence, the electric field strength has a form
\begin{equation}
E(r,t)=\frac{E_o}{r}e^{-H_o (t-t_s)} \exp \left[ \imath \phi (r,t) \right]
\end{equation}
 where $\phi(x,t)$ is the phase in (20).
The intensity of this wave is
\begin{equation}
  I(r,t)=\frac{c(t) \varepsilon (t)}{4 \pi} EE^* =\frac{c_o \varepsilon_o }{4 \pi} \left( \frac{E_o}{r}\right)^{2} e^{-2H_o (t-t_s)}
\end{equation}
In order to estimate decreasing intensity with distance and look-back time
let us use of (16) and (22) at  $t=t_o$  with inserting the parameters of our
epoch $c(t_o),\varepsilon(t_o)$ and $L(\tau_o)=r(t_o)$    into (25).
Following the substitutions  we have
\begin{equation}
 I(L,\tau_o)=I_o \left( \frac{H_o}{c_o} \cdot \frac{e^{-H_o\tau_o}}{e^{H_o\tau_o}-1} \right)^2
\end{equation}
where $I_o =\frac{c(t_s) \varepsilon (t_s) E_{o}^2}{4 \pi}= \frac{c_o \varepsilon_o) E_{o}^2}{4 \pi} $. \\
Let us compare this one with  the intensity of this wave taken at some previous
point of optical path from the source $I(r_1,\tau_1)$ where $r_1=r(\tau_1)$
and $\tau_1=t_1 -t_s$. The distance $r(\tau_1)$ is chosen much shorter  than
$L(\tau_o)$  but long enough  to consider the galaxy as a point source.
With aim to compare we derive the relation $\beta=I(L,\tau_o)/I(r_1 ,\tau_1)$
\begin{equation}
\beta=\left( \frac{H_o \tau_1 e^{-H_o \tau_o}}{e^{H_o \tau_o}-1} \right)^2
\end{equation}
where we have   taken   $\tau_1 << H_o^{-1}$ into account  at series expansion
of the exponent.
According to current data, the Hubble's constant is in the range of $60-140 km s^{-1} Mpc^{-1}$.
Making  use of  $H_o =100 km s^{-1} Mpc^{-1} $ ,i.e. $ 2,9 \cdot 10^{-18} s^{-1}$~, and
$\tau_1=5\cdot 10^6$ years (at $r_1 =100d$  where $d$  is a size of the typical
galaxy equaled of $50 kps$)  to estimate  $\beta (\tau_o)$  and $Z(\tau_o)$ for
 different look-back times we obtain the following table\\
  where $\beta=(\tau_1/\tau_o)^2$ for the situation
   with constant light velocity\\
   
 Table
\begin{table}
\begin{tabular}{|c|c|c|c|}
\hline
$\tau_{o}$  (look-back time) & $Z$  (redshift)  & $\beta$ (at variable $c$)  &  $\beta$ (at constant $c$)  \\
\hline
$H_{o}^{-1}=1.08 \cdot 10^{10} years  $  &  1.8 & $1.5\cdot 10^{-11}$  &
$2\cdot 10^{-7}$ \\
2$H_{o}^{-1}$ &  6.8 &  $1.2\cdot 10^{-13}$ & $5\cdot 10^{-8}$ \\
3$H_{o}^{-1}$ & 21   & $2.1\cdot 10^{-15}$  & $2.2\cdot 10^{-8}$\\
4$H_{o}^{-1}$ & 60   &  $3\cdot 10^{-17}$   & $1.2\cdot 10^{-8}$ \\
5$H_{o}^{-1}$ & 170  & $4.5\cdot 10^{-19}$  & $8\cdot 10^{-9}$  \\
6$H_{o}^{-1}$ & 480  & $7\cdot 10^{-21}$    & $5\cdot 10^{-9}$  \\
\hline
\end{tabular}
\end{table}

   It is seen from here that intensity of light is falling more abruptly in the
case of the variable light velocity compared to the constant light velocity situation.
This appears to  result in very strong restriction of the visible horizon  to
make measuring of $Z$ more than 6 practically impossible at sensitivity of
up-to-day equipment.

\section{Conclusion}
A new model of cosmological redshifts developed on the base
of classical electrodynamics and experimental Hubble's law is discussed in
this paper.In distinction from the usual wave equation the modified one (3)
obtained here has the third term taking into consideration interaction the
light with physical vacuum as a real matter.The light  was concluded from solving
this equation to propagate in vacuum with a constant wavelength
shifted initially (11) and variable velocity (16) caused by gradual changing
of permittivity and permeability of physical vacuum (a relative rate
of about $10^{-10}$ per year).Actually,for very long time of travel of the
light in space from a quasar to the Earth,the wavelength of a terrestrial
source is being shifted due to evolution of the Universe resulting in fractional
redshifts.For this same reason the frequency of the traveling light perceived by
the observer on the Earth is a function of time \\(18 ) to be different from
the frequency $\omega_o$ specified by the energy transition which remains constant
in time at any point of the Universe for any atom in question.
In distinction from the tired-light model,decreasing of the amplitude of the
electric field strenth (19) during the travel trough space is due to not
absorption.From this EVM such a behavior is a result of time evolution of
physical vacuum.
 This model offers novel explanations not only for the redshift origin but
also for several other observed features of reality,for instance,
Olber's paradox and limitation of Z.
\section{Appendix}
Consider the usual wave equation \cite{BW} with stationary boundary conditions
\begin{displaymath}
 \frac{\partial^2 E}{\partial x^2}  -  \left( \frac{\varepsilon \mu}{c_o ^2}\right)\frac{\partial^2 E}{\partial t^2}=0
\end{displaymath}
at  $E_{o} , \omega_{o}, c_o, \varepsilon_{o}$  and  $\mu$ - constant \\
         and $E(0,t)=E_{o} \exp \imath \omega_{o}t$
to study the propagation of light at {\em constant} velocity in any
non-conductors,including vacuum.
In order to discuss the tired-light model let us seek a solution of this
 equation as a quasi-periodic function with variable phase  and x-dependent
amplitude
\begin{displaymath}
 E=b(x) \exp \imath  \phi (x,t)
\end{displaymath}
Inserting this into the equation and separating real and imaginary
parts we obtain two following equations
\begin{eqnarray*}
b'' - b (\phi') ^2 +b \left( \frac{\varepsilon \mu}{c_o^2} \right) (\dot\phi)^2 =0 \\
2b'\phi ' +b\phi'' =0
\end{eqnarray*}
where $\dot \phi=\omega_o=const.$\\
 Taking  $\phi'=k=\frac{2\pi}{\lambda}$ into account the second one can be
rewritten  as follows
\begin{displaymath}
 2b'\lambda - b\lambda' =0
\end{displaymath}
It follows from here that $b/\lambda=2b'/\lambda'$.Because the left side
is always positive one the x- derivative signs for $b$ and $\lambda$
should be this same.It means that either we have redshifts ($\lambda'> 0$)
with increasing of the amplitude (that is absurd) or decreasing
amplitude ($b'< 0$) and violetshift ($\lambda'< 0$,in no compliance with experiments).\\
Thus the situation of tired-light model ($\lambda'> 0$ with $b'< 0$) is not
acceptable for the usual wave equation with stationary conditions.


\begin{thebibliography}{14}
\bibitem{hu} E.P.Hubble, Proc. Nat. Acad. Sci. 15, 168 (1929).\\
\bibitem{tu} R.Brent Tully, Astrophysical Journal,303,25(1986). \\ R.Brent
Tully,J.R.Fisher,Atlas of Nearby Galaxies,(Cambridge:Cambridge University Press,1987)\\
\bibitem{ler} E.J.Lerner, The Big Bang never happened,(Simon \& Schuster Ltd,London,1992).\\
\bibitem{di} P.A.M. Dirac, Nature 139,323 (1937).\\
\bibitem{PV} A. Potekhin,D. Varshalovich, Astronomy and Astrophysics Supplement 104,89 (1994).\\
\bibitem{ar} H. Arp,Progress in New Cosmologies,(Plenum Press, New York,1,1993).\\
\bibitem{arp} H. Arp,Quasars, Redshifts and Controversies,Interstellar Media,(Berkeley,1987).\\
\bibitem{DeB} L.De Broglie,Cahiers de Physique,16,425 (1962).\\
\bibitem{S} E.I.Shtyrkov,Gal.Electrodynamics,3,66 (1992).\\
\bibitem{Sh} E.I.Shtyrkov,Progress in New Cosmologies, Plenum  Press, New  York,327 (1993).\\
\bibitem{Sht} E.I.Shtyrkov,Gal. Electrodynamics,8,3,57 (1997).\\
\bibitem{MD} A.Montgomery,L.Dolphin,Gal.Electrodynamics 5,93 (1993).\\
\bibitem{Ol} H.W.M.Olbers,Edinburg New Philosophical Journal,1,141 (1826).\\
\bibitem{BW} M.Born,E.Wolf,Principles of optics,(Pergamon Press,New York,1964)\\
\end{thebibliography}
\end{document}